\documentclass[screen]{acmart} %
\usepackage{graphicx}
\usepackage{soul}
\usepackage{tabularx}
\usepackage{titlesec}

\usepackage{xcolor}
\usepackage[normalem]{ulem} 

\newif\ifshowrevisions
\showrevisionsfalse 

\ifshowrevisions
  \newcommand{\rev}[1]{\textcolor{blue}{#1}}
  \newcommand{\revdel}[1]{\textcolor{red}{\sout{#1}}}
\else
  \newcommand{\rev}[1]{#1}
  \newcommand{\revdel}[1]{}
\fi

\AtBeginDocument{%
  \providecommand\BibTeX{{%
    \normalfont B\kern-0.5em{\scshape i\kern-0.25em b}\kern-0.8em\TeX}}}

\setcopyright{acmcopyright}
\copyrightyear{2018}
\acmYear{2018}
\acmDOI{XXXXXXX.XXXXXXX}

\acmConference[Conference acronym 'XX]{Make sure to enter the correct
  conference title from your rights confirmation emai}{June 03--05,
  2018}{Woodstock, NY}
\acmPrice{15.00}
\acmISBN{978-1-4503-XXXX-X/18/06}

\newcommand{\elide}[1]{\textelp{}} 
 
\begin{document}

\title{“I’m Constantly Getting Comments Like, ‘Oh, You’re Blind. You’re Like the Only Woman That I Stand a Chance With.’”: A Study of Blind TikTokers’ Intersectional Experiences of Gender and Sexuality}

\author{Yao Lyu}
\orcid{0000-0003-3962-4868}
\affiliation{%
  \institution{University of Michigan}
  \city{Ann Arbor}
  \state{Michigan}
  \country{USA}
}
\email{yaolyu@umich.edu}

\author{Jessica Shen}
\affiliation{%
  \institution{University of Michigan}
  \city{Ann Arbor}
  \state{Michigan}
  \country{USA}
}
\email{jesishen@umich.edu}

\author{Alina Faisal}
\orcid{0009-0000-9495-8057}
\affiliation{%
  \institution{University of Michigan}
  \city{Ann Arbor}
  \state{Michigan}
  \country{USA}
}
\email{alinafai@umich.edu}

\author{John M. Carroll}
\orcid{0000-0001-5189-337X}
\affiliation{%
  \institution{Pennsylvania State University}
  \city{University Park}
  \state{Pennsylvania}
  \country{USA}
}
\email{jmcarroll@psu.edu}

\renewcommand{\shortauthors}{}


\begin{abstract}

Social media platforms are important venues for identity expression, and the Human-Computer Interaction community has been paying growing attention to how marginalized groups express their identities on these platforms. Joining the emerging literature on intersectional experiences, we study blind TikTokers (“BlindTokers”) who are also women and/or LGBTQ+. Using interview data from \rev{41} participants, we identify their intersectional experiences as mediated by TikTok’s socio-technical affordances. We argue that BlindTokers’ intersectional marginalization is infrastructural: TikTok’s classification and moderation features interact with social norms in ways that push them aside and distort how they are treated on the platform. We use this infrastructure perspective to understand what these experiences are, how they were formed, and how they become harmful. We further recognize participants’ infrastructuring work to address these problems. This study guides future social media design with accessible creator tools, inclusive identity options, and context-aware moderation developed in partnership with communities.
\end{abstract}




\maketitle

\section{Introduction}

Human-Computer Interaction (HCI) and Computer-Supported Cooperative Work (CSCW) communities have long paid attention to understanding user experiences on social media platforms. Specifically, researchers have documented how users engage with these platforms for identity work and community building. Because of the emphasis on the social implications of these platforms, recent research has shifted away from technical aspects and toward social impacts, particularly those affecting marginalized populations \cite{simpson_for_2021,lyu_i_2024-1}. TikTok has become increasingly popular over the last few years due to viral trends and the opportunities it offers for marginalized populations to connect, build community, and gain visibility. However, TikTok's algorithmic system has been reported to force marginalized users to adapt their behaviors and adopt strategies to protect their content \cite{lyu_i_2024,simpson_how_2022,biggs_tiktok_2023,duguay_tiktoks_2023}. The "For You" page dictates what content is seen, and often, content by marginalized individuals can be overridden by this algorithmic curation regardless of tags or follower networks \cite{duguay_tiktoks_2023}. According to reports, this algorithmic curation suppresses content by minorities and labels the reasoning as “protecting” these users \cite{kover_discrimination_2019}. Scholars have also found that these moderation practices can inadvertently perpetuate biases and result in further marginalization \cite{ball-burack_differential_2021}. To avoid being shadowbanned, creators have developed and tested their own folk theories about the algorithmic systems and created strategies, including using coded language, self-censoring, and avoiding identity-related vocabulary to protect their content and communities \cite{duguay_tiktoks_2023,devito_adaptive_2021, devito_how_2022}. These strategies, however, produce negative implications, as users are forced to change the way they behave and speak.

Recent studies have increasingly focused on understanding social media through more holistic concepts that foreground marginalized groups and their experiences, especially as social media research now draws heavily from social and critical theory to consider platforms’ potential for liberation \cite{duguay_tiktoks_2023} or domestication experiences toward algorithms \cite{simpson_how_2022}. Though HCI research on social media has extensively documented online marginalization, most studies analyze harms along a single axis of identity (e.g., gender \cite{cui_so_2022}, sexuality \cite{simpson_how_2022}, or disability \cite{lyu_i_2024-1}), even when acknowledging that multiple identities are at play \cite{lyu_i_2024}. As a result, intersectional marginalization—how co-present identities combine to shape experiences—remains underexplored. Among the small body of work that does take intersectionality, especially disability and gender or sexuality, seriously \cite{crawford_complex_2023,crawford_like_2025}, there is still limited theorization of how algorithmic platforms mediate, induce, and reproduce intersectional experiences through ranking and recommendation systems, moderation pipelines, visibility metrics, and identity legibility requirements.

Our study addresses these gaps by examining blind TikTokers (“BlindTokers”) who are women or LGBTQ+. While prior work often groups women and LGBTQ+ users together due to overlapping forms of marginalization \cite{freeman_disturbing_2022,uttarapong_harassment_2021,zytko_online_2023}, we analyze both convergences and nuanced divergences in their experiences, showing where patterns align and where they importantly differ. We guide the study using two research questions:

\begin{itemize}
    \item RQ1: What are the intersectional experiences of Blind Women/LGBTQ+ TikTokers?
    \item RQ2: How do Blind Women/LGBTQ+ TikTokers navigate these challenges?
\end{itemize}

To address these questions, we conducted 41 semi-structured interviews with blind creators who are also sexual or gender minorities. We employed a thematic analysis to examine the data and identify key themes related to our research questions. Participants reported that TikTok’s limited gender disclosure options and content moderation system flagged identifier words like “blind” or “queer” in their content. We found two main strategies creators used to navigate these challenges: internalizing and self-disciplining behaviors, such as intentionally modifying their clothing to avoid being flagged, and resisting with community support by finding groups of individuals with shared identities. We further adopted infrastructure theory to conceptualize intersectional experiences not as isolated incidents but as infrastructural effects—outcomes of classification schemes, data and moderation workflows, engagement-driven visibility logics, and the everyday practices that bind online and offline life.

This study has several contributions to the literature. (1) It documents empirical evidence showing how blind women or LGBTQ+ users encounter intersectional marginalization on TikTok, highlighting nuanced lived experiences shaped by platform design, algorithmic management, gender and sexuality, and visual impairment. (2) It unpacks these intersectional experiences using infrastructure theory to explain how platform arrangements can simultaneously enable expression and connection while also misclassifying identities, amplifying harassment, and restaging offline inequities at scale. (3) It proposes design implications for the future development of platforms. With these contributions, we enrich HCI’s intersectional scholarship and offer a systematic account of how algorithmic infrastructures shape, intensify, and sometimes reconfigure the lived experiences of blind women and LGBTQ+ users on social media.

\section{Related Work}

\subsection{The Blind Community on Social Media}

According to the World Health Organization, by 2023, approximately 2.2 billion people worldwide experienced vision impairment or blindness \cite{world_health_organization_vision_2023}. Prior research has shown that visually impaired people are heavy users of social media \cite{lyu_because_2024,stangl_person_2020}. Social media is part of everyday life for many blind or low vision (BLV) users, allowing them to connect with others in their community while giving them space to share their experiences and navigate their identities \cite{lyu_i_2024-1,alonzo_review_2025}. Social media is used not only for factual information but also for emotional and social engagement, which often involves images, GIFs, and videos \cite{xu_danmua11y_2025}. However, most platforms are designed for sighted users, resulting in visually dense layouts, inconsistently labeled text and buttons, and inaccessible keyboard navigation, all of which make interaction difficult for BLV users \cite{aljedaani_challenges_2025}. Recent studies highlight that BLV users face significant challenges in online shopping due to inadequate image descriptions \cite{wang_revamp_2021}, touchscreen-based image exploration because they cannot survey clustered areas or locate regions \cite{nair_imageassist_2023}, and audio- or feedback-related accessibility issues, in addition to insufficient customization, personalization, and persistent algorithmic biases \cite{aljedaani_challenges_2025}. Although BLV users adopt creative strategies and assistive technologies to navigate inaccessible platforms, these solutions require substantial effort.

\rev{Recently, HCI literature has shown growing interest in making online content more accessible to BLV users by advocating for assistive technologies \cite{stangl_person_2020,stangl_going_2021,gleason_addressing_2019}. BLV users frequently encounter challenges engaging with photos \cite{dos_santos_marques_audio_2017}, videos \cite{jun_exploring_2021}, emojis \cite{tigwell_emoji_2020}, and memes \cite{gleason_making_2019}, which often lack alt-text descriptions.} For example, locating academic information in research papers is challenging for BLV users due to dense writing, inconsistent structure, and poorly labeled figures and tables \cite{park_exploring_2022}. Hyperlinks to citations are often one-directional, making it difficult to return to the main text \cite{park_exploring_2022}. BLV users also struggle to access digital news platforms due to poor use of headings and landmarks, non-intuitive layouts for navigating articles, and news videos that rely heavily on on-screen text and music without spoken descriptions, discouraging exploration \cite{mowar_breaking_2024}. Some visual design barriers can also mislead users of accessibility technologies. For instance, Lewis et al. \cite{lewis_inaccessible_2025} describe how screen readers label unlabeled or hidden interface elements simply as “button” or nothing at all, leading to accidental subscriptions or purchases. “Unsubscribe” buttons are also often buried deep within interface layers, resulting in poor keyboard navigation \cite{lewis_inaccessible_2025}. To mitigate such challenges, recent work explores how Generative AI can enhance or automate the generation of alternative text for images, summarize inaccessible visual data such as charts, and provide auditory descriptions of formatting and layout issues \cite{perera_sky_2025}. For instance, Li et al. \cite{li_recipe_2024} show that participants following recipes with screen readers struggle due to long, unstructured steps and cluttered websites filled with advertisements that interrupt the reading experience \cite{li_recipe_2024}. Researchers have also compared platform design choices across Twitter and Facebook. Twitter initially offered more accessible design features, as screen readers could easily read aloud the platform’s primarily textual content \cite{gleason_twitter_2020}. However, as Twitter evolved and users increasingly posted images and visual content, the platform became more cluttered and less accessible \cite{gleason_twitter_2020}. By contrast, Facebook was designed initially for sighted users, creating substantial usability and accessibility issues for blind users when navigating via screen readers \cite{pakdeechote_new_2012}. Its multilayered components, such as headers and sidebars, further hinder navigation \cite{pakdeechote_new_2012}.

Blind users rely on social media networks to form online communities where they can share lived experiences, engage in discussions, pursue professional development, and seek entertainment \cite{lyu_because_2024}. Such connections help them feel heard and provide a sense of belonging, while also motivating them to support other blind creators’ content and defend friends against harassment or trolling \cite{lyu_because_2024}. For example, Li \cite{li_it_2022} describes how social media enables blind people to connect with others to learn, share, support, and exchange information and skills. Blind creators on YouTube produce video tutorials—such as makeup guides—that serve as resources for other BLV users, reducing dependence on content produced by sighted creators and fostering mutual support \cite{li_it_2022}. Another study by Lyu et al. \cite{lyu_i_2024} highlights TikTok as an important platform for blind users to engage in informative, educational, and friendly exchanges with sighted people. Comments from sighted viewers, often driven by curiosity or unfamiliarity with blindness, create opportunities for meaningful interaction \cite{lyu_i_2024}. However, blind users also face harassment, trolling, and algorithmic moderation on platforms like TikTok, where they encounter accusations of “faking blindness” and judgmental comments \cite{lyu_i_2024}. The study also reveals how their content is frequently suppressed because it does not align with platform norms or recommendation logic \cite{lyu_i_2024}. These experiences illustrate how social media simultaneously functions as a space of empowerment and a site of exclusion for blind communities.

Although prior research has documented challenges facing blind and visually impaired users on social media, significant gaps remain in understanding the experiences of BLV users who also belong to gender and sexual minority groups. The experiences of individuals with multiple marginalized identities provide critical insight into how they navigate social platforms and underscore the need for better platform designs that support diverse and inclusive online communities. In this article, we focus on blind creators who identify as gender and/or sexual minorities. By foregrounding their experiences, we contribute to understanding how accessibility barriers and gender/sexuality biases intersect, revealing the complex challenges faced by marginalized groups on social media platforms.

\subsection{Gender and Sexual Minorities on Social Media}

We also review literature on gender and sexual minority communities on social media. Gender is a multifaceted, complex concept that consists of multiple aspects, including physiological or bodily attributes (sex), gender identity or self-defined gender, legal gender, and social gender, in terms of norm-related behaviors and expressions conventionally labeled as feminine or masculine \cite{lindqvist_what_2021}. \rev{Sexual minorities, like LGBTQ+, encompass a range of gender and sexual identities and expressions that diverge from societal norms, including but not limited to lesbian, gay, bisexual, and transgender individuals \cite{cochat_costa_rodrigues_sexual_2017}.} Social media has become a space with the potential for LGBTQ+ individuals to find joy and pride in their identities, achieve representation, defy stereotypes, and engage in activism \cite{kender_social_2025,liu_navigating_2025,amirkhani_society_2025}, as well as explore identity \cite{cullen_not_2025}, obtain peer support, and access (sub)cultural information \cite{byron_hey_2019}. For example, Tumblr has become a social space where users share accounts of experiences with mental health, physical health, gender, and sexuality because it centers on a culture of disclosure, anonymous support, and feeling heard. This work highlights key aspects of an online queer space that Tumblr provides, allowing LGBTQ+ individuals to experience multiplicity, fluidity, and ambiguity in a comfortable environment, while helping them learn to cope with ongoing discomfort \cite{byron_hey_2019,haimson_tumblr_2021}.

There has been increasing HCI research on understanding the challenges women or LGBTQ+ people face on social media platforms and the strategies they use to access resources, explore identities, and find reliability \cite{lucero_safe_2017}. One central concern for LGBTQ+ individuals is having the autonomy to control when and how to be “out” on social media across different audiences—particularly the ability to separate their offline personal lives from their online presence within LGBTQ+ communities—even though many platforms still fail to support “a selective and intentional process of LGBTQ+ self-presentation” \cite{carrasco_queer_2018}. Because platforms do not offer sufficient control over the visibility of identity presentation for different audiences, LGBTQ+ users often turn to strategies such as self-censorship \cite{carrasco_queer_2018}. Another common strategy is managing identities across multiple accounts: one for LGBTQ+ friends and another for friends and family. This allows them to lead a “double life” to manage separate social circles, “avoid potential conflict or discrimination,” and protect relationships and identities from non-LGBTQ+ acquaintances who may not share the same perspectives or beliefs \cite{zhu_challenges_2025}. For example, research shows that Facebook’s audience management failures—and its prohibition of multiple accounts—prevented many LGBTQ+ participants from both satisfying their families and using the platform to connect meaningfully with friends \cite{carrasco_queer_2018}. Through social media, blogs, and online forums, LGBTQ+ individuals gain spaces to “connect, share experiences, join community,” and feel seen and accepted \cite{bhandari_analysis_2024}. Engagement in online LGBTQ+ communities has also inspired some individuals’ coming-out processes, increasing their confidence in their identities. Due to its scale, social media has become a powerful tool for queer activism—used “to advocate rights, to educate others, and to mobilize support during pivotal social movements.” Digital advocacy efforts such as \#LoveWins and \#TransRightsAreHumanRights have been widely used to promote personal stories and create visibility and awareness around LGBTQ+ issues \cite{bhandari_analysis_2024}. Yet platform designs can still exclude users by design. Fadrigon’s study of user account creation processes shows how LGBTQ+ individuals face exclusion when noninclusive practices fail to represent the fluidity and complexity of their identities. As a result, participants felt uncomfortable and hesitant to share personal information due to distrust toward platforms that do not permit nuanced or optional representation of gender and sexuality \cite{fadrigon_playbook_2024}.

The misalignment between platform values and user values has heightened risks and marginalization that queer users already face, pushing some to avoid certain platforms and seek media that provide critical opportunities for self-identity. Social platforms have their own goals, such as “platform growth, boosting content engagement, and avoiding legal trouble” \cite{devito_values_2021,gillespie_politics_2010}. Online marginalized communities, however, rely on these spaces for identity development and informational and social support. To work around platform limitations, these communities must expend substantial effort to remain supported and sustainable. Recent work therefore investigates alternative media that can offer critical opportunities for self-identity for LGBTQ+ individuals—one of which is virtual reality. Virtual reality has become increasingly popular among queer users \cite{li_we_2023,freeman_acting_2022}. VR’s unique affordances, which allow users to connect their physical bodies and engage in self-preservation within digital spaces, reshape how people communicate, connect, and interact. In comparison, traditional platforms often rely on following, subscribing, or donating behaviors that limit emotional and social cues \cite{li_we_2023}. Social virtual reality has grown in popularity due to its potential to support the “transformative nature of embodied visibility” and offer queer users a “safe sanctuary that is disconnected from their offline lives” \cite{freeman_acting_2022}.

In this paper, we join research on women and LGBTQ+ communities on social media. We extend this work by drawing attention to a more specific group of social media users who experience marginalization not only on the basis of gender or sexuality, but also on the basis of disability status. We use this case to further unpack the nuanced lived experiences of intersectional marginalization and revisit identity construction in online spaces.

\subsection{Intersectional Experiences and Social Media Infrastructure}

Our work also revolves around the connections among intersectionality, platform governance, and infrastructure studies. This review defines and explains the key concepts we build upon and identifies the gaps our study aims to address. Intersectionality describes overlapping power dynamics—including sexism, ableism, racism, and cisheteronormativity—that shape lived experiences \cite{crenshaw_mapping_1991}. In HCI research, intersectionality has been used to critique uniform design solutions and center marginalized users in design outcomes and methods \cite{schlesinger_intersectional_2017,costanza-chock_design_2020,dignazio_personal_2020}. Kafer and Hamraie argue that access should be understood as both socio-technological and political, rather than purely ergonomic, as highlighted by disability and queer scholarship \cite{kafer_feminist_2013, hamraie_2013_designing_2017,hamraie_building_2017}. Bear Don’t Walk IV et al. \cite{bear_dont_walk_opportunities_2024} further argue that intersectionality studies should consider systemic privilege, interrelatedness, nuance within participant groups, problems of categorization, and the need for research to inform meaningful change. Together, these frameworks are essential for understanding how blind women/LGBTQ+ creators navigate content and expression on TikTok.

Infrastructure studies highlight how categories, standards, and maintenance shape everyday practice. A key property emphasized is that infrastructure—anchored by its installed base—is often invisible in use but becomes visible when it breaks, revealing its politics as well \cite{star_steps_1996,edwards_infrastructure_2010,larkin_politics_2013}. Bowker and Star emphasize that classification is a form of infrastructural work \cite{bowker_sorting_1999}. Categories make people and practices legible; however, they also misclassify, creating residuals and torque. Torque refers to the friction that arises when biographies clash with fixed categories \cite{singh_margins_2017}. Imbrication describes how social norms and technical features intertwine; thus, scholars have begun treating platforms as infrastructure \cite{singh_margins_2017,plantin_wechat_2019}, showing how recommendation, moderation, and reporting systems shape visibility patterns and user engagement. Scholars in HCI extend platform studies by relating them to maintenance, repair, and infrastructure—namely, the ongoing work done by users and communities to keep vulnerable systems usable \cite{lyu_systematic_2025}. Platform studies further emphasize how moderation and algorithmic recommendation systems apply practices such as gender presentation cues and reclaimed identity terms to elevate or suppress identity expression \cite{gillespie_politics_2010,myers_west_censored_2018,matias_civilservant_2018,noble_algorithms_2018,benjamin_race_2023}. Within this context, platforms implement governance through keyword filtering, visual and modesty heuristics, and reporting dynamics—all of which influence how identities circulate and become visible. Research on blind women creators highlights visible gaps and disparities between creative accessibility and accessibility for producing content, spanning labeling, authoring, and editing. We analyze labeling, authoring, editing, accessibility, and intersectional identity fields together.

We integrate and bridge insights from infrastructure and intersectionality through a pipeline model. The pipeline proceeds sequentially: classification leads to imbrication, then residuality, and finally torque, illustrating how intersectional experiences emerge on algorithmic platforms. These harms arise both from technical features of the platform and from infrastructural violence—epistemic, emotional, and interpersonal \cite{rodgers_infrastructural_2012, li_there_2021}. Our work builds on prior research by specifying accessibility requirements (e.g., effect descriptors and text-to-speech–legible pronoun fields), reconceptualizing creator tactics (e.g., coded words, mutual-aid groups, wardrobe rules) as infrastructuring practices, and proposing system-level repairs (e.g., context-aware moderation, anti-brigading protections, community handoff scaffolds) that reduce burdens while addressing infrastructural causes. Together, this review clarifies the theoretical foundations of our work and highlights the gaps our study aims to address.

\section{Methods}

\subsection{Study Overview}
In this section, we will articulate the methods, including data collection (i.e., semi-structured interviews) and data analysis methods (i.e., thematic analysis). This study is part of an ongoing research project on how people with visual impairments experience social media platforms, specifically on TikTok. \rev{This study was approved by the Institutional Review Board (IRB) at the first author’s doctoral institution.} At the beginning, we created an official TikTok account using the first author's real name and affiliation information; we kept posting information about the first author and the project. This step was to build trust with the blind community on TikTok. Then we started searching for users with visual impairments using "blind", "cane user", or other variants; we followed all the users we searched and actively engaged with their content creation, like videos and live streams; we left positive feedback to support the users with visual impairments. After we built trust with the creators, we encouraged them to follow us back so we could provide more detailed information about our project and invite them to participate. \rev{This sampling process was purposive and availability-based. We first identified BlindTokers who could potentially be participants, then reached out to them. We acknowledge that this approach has limitations, as we may not have reached blind users who did not include keywords such as "blind" or "visual impairment" in their bios, profiles, or content descriptions.}

\rev{To ensure accessibility, we conducted interviews according to participants' preferences. All interviews were held via Zoom, and demographic information was collected orally. Because interviews were fully remote, we obtained oral consent at the beginning of each session. We first read the project information and consent form aloud, confirmed participants' consent, and then proceeded with the interview. With participants' permission, we also recorded audio for data analysis.}



\subsection{Data Collection: From Overall Accessibility Issues to Intersectional Experiences}

As mentioned previously, the current study is a part of the ongoing project aimed at exploring the overall perceptions of TikTokers with visual impairments on accessibility issues of TikTok. We interviewed all participants via Zoom meetings or phone calls. The interviews took place between 2022 summer and 2023 spring. All interviews were recorded and later transcribed into text files. For the early participants, our interview protocol focused on their everyday social lives on TikTok, their overall usage of TikTok, and the accessibility challenges they encountered on TikTok. The example questions included: "What content do you consume most on TikTok?" "Which content creators do you follow, and why?" "What accessibility challenges do you experience when using TikTok?" "How do you communicate with other TikTokers about the accessibility issues?"

As the study progressed, a pattern emerged among participants who identified as women or LGBTQ+. They disclosed their marginalized experiences not only about their disability but also about their gender/sexuality. We noticed this pattern and modified our interview protocol to address this intersectional topic. In the later interviews, we also asked questions about their experiences with their gender/sexuality on TikTok, such as "How do you disclose your gender/sexuality on TikTok?" "How do you communicate with other TikTokers about your gender/sexuality identities?", especially, we also asked about the intersectionality experiences, like "How do you describe your experiences of navigating TikTok as a visually impaired LGBTQ+ individual?", we also asked about TikTok's role, like "How does the TikTok platform mediate all the mentioned experiences?" At the end of the study, 41 participants provided insightful answers to these interview questions.

Our participants included 41 BlindTokers with their age range spanning from 18 to 62 (average 33). Among all participants, 34 participants identified as female, 3 participants identified as male, and 4 participants were nonbinary. \rev{Regarding sexuality, most participants (34) did not disclose this information. Five told us they were LGBTQ+ without specifying a category, while one identified as lesbian and another as asexual.} All the gender and sexuality information was documented as it was self-reported by participants. Additionally, 33 participants were legally blind, and 5 participants reported low vision, with 3 participants being totally blind. \rev{Low vision refers to vision loss that cannot be increased with equipment or surgery but still allows for some functional use of sight in daily activities; legally blind refers to severe visual impairment that meets clinical criteria and does not allow functional vision, though some light perception may remain; totally blind refers to the complete absence of light perception. We created this visual impairment categorization based on prior HCI research on people with visual impairments \cite{lyu_because_2024,rong_it_2022,cheema_describe_2025}.} The participants’ education varied, as 13 participants had Associate degrees, 8 graduated from High School (H.S.), 9 had bachelor's, and only 3 had General Educational Development (GED) diplomas. 


\subsection{Data Analysis: Six Step Thematic Analysis}

\rev{We conducted a reflexive thematic analysis \cite{braun_reflecting_2019,braun_using_2006} of the interview data. Our analytic process followed an abductive approach, beginning with inductive coding to capture a wide range of possible patterns. As themes emerged, we refined our analysis using a deductive lens informed by infrastructure theory, which we identified as a strong conceptual fit with the patterns generated through open coding.}

Specifically, this method is a six-step process: (1) We read through all the interview transcripts to familiarize ourselves with the data and gain an understanding of the participants' experiences on TikTok and the platform's role. We noted down important quotes on unique experiences and overall shared insights during weekly meetings.(2) After gaining an overall understanding, we identified and coded all the quotes that were noted down that described these unique interactions that were relevant to an identity (gender, sexuality, and disability), the individuals' role on TikTok, as well as TikTok's effect on their identity (examples of codes are shown in Table-\ref{tab:code}). (3) Next, we merged the codes and grouped them into themes and sub-themes. (4) We refined the themes based on their conceptual coherence and theoretical implications. (5) We reviewed the themes and noticed that the participants had a variety of interactions on the platform, and could be categorized as "Re-Encountering Offline Intersectional Marginalization", "Encountering Platform-induced Intersectional Marginalization", and "Dealing with Intersectional Marginalization on TikTok." These included the interactions, challenges, and strategies that intersectional marginalization TikTok creators came across or implemented on TikTok. Each theme and sub-theme named is ensured to be representative, exhaustive, and mutually exclusive. (6) Lastly, we reported all the themes and sub-themes with examples and elaborations in the findings section. This analysis combines inductive coding and deductive coding 


\begin{table}[htbp]
    \centering
    \caption{Example of Code of Intersectional Experiences}
    \renewcommand{\arraystretch}{2}
    \begin{tabularx}{\textwidth}{|p{\dimexpr 0.2\textwidth-2\tabcolsep} | p{\dimexpr 0.3\textwidth-2\tabcolsep}|p{\dimexpr 0.5\textwidth-2\tabcolsep}|}
        \hline
        \multicolumn{1}{|c|}{\textbf{Code}}
         & \multicolumn{1}{c|}{\textbf{\textbf{Definition}}} & \multicolumn{1}{c|}{\textbf{\textbf{Quote}}} \\
        \hline
        \textbf{Encounter Sexual Harassment} & TikTokers received uninvited offers or questions about sexual comments & "\textit{...first questions they start saying is, Hey, pretty and beautiful. And then it'd be like, Alright, where are you from? Are you single or married?}[T14]" \\

        \textbf{Get/Provide Group Support} & BlindTokers created support groups to help each other out &  "\textit{...she posted a video as that she was gonna do a support group on WhatsApp for any moms with visual impairment or blindness...} [T13]" \\
        
        \textbf{Denial on Intersectional Identity} & Blind and LGBTQ+ TikTokers got questioned about their intersectional identity & "\textit{...she was like you cannot be blind and LGBTQ at the same time. I'm like, wait, what?} [T09]"  \\
        \hline
    \end{tabularx}
    \label{tab:code}
\end{table}

\subsection{Author Positionality}
We acknowledge that our identities have shaped this study. We are sighted, non-disabled researchers; some of us are cisgender and straight-identifying; none of us are members of LGBTQ+ communities. Although we are not active participants in LGBTQ+ or blind communities, we approach this work as learners and collaborators rather than arbiters of experience. From a social constructionism perspective \cite{sparkes_narrative_2008}, we recognize that our understandings are shaped by academic training, prior research, and—most importantly—the lived experiences shared by BlindTok participants. We also recognize our academic privileges (access to resources, platforms, and audiences) and the unequal distribution of these privileges.

To mitigate bias, we practiced sustained reflexivity and embraced the notion of contamination as collaboration \cite{tsing_mushroom_2015}, working across differences with humility. Following the principle of nothing about us without us \cite{charlton_nothing_2004} and similar guidance \cite{callus_being_2019}, we prioritized listening to participants’ narratives, conducted member-checking when possible, consulted accessibility experts and disability advocates, documented analytic decisions, and treated disagreement as data rather than error. Our prior collaborations with blind participants (e.g., studies of smartphone health applications) sensitized us to how small design choices can exclude and how disability intersects with socioeconomic status and education. Situated within academic structures that have historically marginalized disabled and gender/sexual minority communities, we commit to co-creating knowledge, challenging ableism and heteronormativity, and translating findings into inclusive research practices and outcomes that are usable and equitable.

\section{Findings} 
In this section, we will answer the two research questions with three themes. The first two themes, including "Re-Encountering Offline Intersectional Marginalization" and "Encountering Platform-induced Intersectional Marginalization", answer the first research question; the last theme, "Dealing with Intersectional Marginalization on TikTok," answers the second research question. Details of the themes are reported below.

\subsection{Re-Encountering Offline Intersectional Marginalization}

Our findings highlight that TikTok often produces the same forms of marginalization that participants are already facing offline. Such patterns emerged at the crossroads of gender, sexuality, and blindness, shaping the way people interact, perceive, and engage with participants. Our analysis focuses on two key patterns: (a) ongoing cultural rejection and (b) intensified, heightened vulnerability. Participants identified these as regular, ordinary, rather than exceptional and unusual dynamics of being on TikTok. Participants were already familiar with the dynamics from offline life, but TikTok’s features like comment threads, duets, and open DMs amplified them at a greater scale. 

\subsubsection{Navigating Extended Cultural Rejection}
All participants (N=41) faced the issue of extended cultural rejection, according to which the audience questioned the legitimacy of their identities. Several comments or posts on TikTok often associate blindness with an actual condition of vision loss and LGBTQ+ or queer as something fake, highlighting that an individual cannot be both at the same time. T41 highlights that people on TikTok point towards identity misrecognition, denying the idea that queer and blind identities exist together:

\begin{quote}
    \textit{A lot of people... they're like, "You're like, how many possible things? You can't be both. You're making it all up to get attention." I'm sure people in real life do experience this, but as far as TikTok goes, this is kind of a childish, chronically online idea that has really made waves. People love to argue whether sexuality is a choice or not. But as far as blindness goes, it's a very blatantly medical condition. None of us could possibly choose to have cataracts.} [T41]
\end{quote}

Commenters insist that blindness is “real” as it is a medical issue, while queerness is a “choice” and equivalent to seeking attention. Such strict boundaries make people’s overlapping or multiple personalities seem fake and do not show them as binary individuals. Similarly, ableism and queer-phobia reinforce the idea that a blind individual is legible and sexuality is optional. TikTok’s comment section and algorithm keep this debate alive and push controversial content in the comments to the top, so the denial related to individuals' identities resurfaces and spreads widely. Therefore, due to TikTok’s algorithm, it repeats the same issue or judgment to people offline as well, making it a part of users’ online experience. T41 shows that people easily dismiss the acceptance of multiple identities by using the contrast between “medical factual logic” and “individual choice” while acknowledging blindness and dismissing queerness. Such a way of discussing the issue with medical reasoning and logic for making blindness the only “real” identity, while dismissing queerness and considering it as fake, eliminates the reality that both identities can exist together. Ableism and anti-queer bias intersect each other as blindness is considered factual medical information, whereas queerness is portrayed as illegitimate. Participants had gender-related doubts regarding their ability to parent and also caregiving, which reflected their everyday-to-day stereotypes.  


In addition, participants shared that they encountered questions on their capabilities, such as parenting as women. T21 highlights how kids are impacted when people stare at the cane their mothers carry. 

\begin{quote}
    \textit{And I think there was another (video of a blind mom), she was interviewing her kids, what does it feel like to have a blind mom. And I think the thing that bothers the kids the most, because they can see, is that people stare when their mom uses the cane...For a while, I was afraid to have kids, because I wasn't sure...when I'm out and with just my kids. I would be afraid that I would lose them.}[T21]
\end{quote}

It shows that the gaze of the public on mobility aids allows strangers to point out the parental ability to take care of their kids, creating a stigma and spreading fear in society about whether someone is fit to be a parent or not. The expectations of being an ideal and a “good mother” fuse with stereotypes, showing that blind people are incapable, making parenting a public exam. Additionally, this criticism grows online in comment sections by the public, when videos of blind mothers appear, showing how the pattern continues on social media as well. Here, the way of sighted parenting turns into a public trial for the blind mothers, as the expectations of motherhood combine with ableist views. Fear of judgment and self-doubt arise as people stare at the mobility aids, which permits strangers to question the perfection of motherhood by mixing it with stereotypes of blindness. The comment culture on social media magnifies the offline biases and amplifies the pattern of doubting identity, including ability.

\subsubsection{Experiencing Amplified Vulnerability} 

Another sub-theme shows that TikTok plays a role in magnifying participants' faced vulnerabilities by making them more exposed to strangers' harassment. Nearly half of the participants (N=18) emphasized feeling unsafe on TikTok, as it was a toxic platform for discussion on disability, gender, and sexuality related topics, and strangers could easily reach out. T04 recounted receiving inappropriate and invasive sexual questions alongside age-specific suggestions and propositions through direct messages: 


\begin{quote}
    \textit{Somebody, they ask questions about toileting, "how do you know when you are finished, make sure it's cleaned?"...and also you get loads of private messages of people looking for mature women over the age 35: “here's my phone number, call me” I think I get about ten of those every single day.} [T04]
\end{quote}

Therefore, it indicates that TikTok does not replace the already existing offline risks; rather, it amplifies them. Instead of decreasing vulnerability, TikTok magnifies and signifies the online digital attention by attracting strangers to intrude and cross the boundaries of connection. Participants often explained harassment through sexual, inappropriate comments and unwanted direct messages that fetishized blindness and objectified gender norms. This shows how blindness is used as an excuse for harassment, which works similarly to sexually monitoring and targeting women through age-based propositions. It also demonstrates that blindness is fetishized, while magnifying the vulnerability of people’s behavior on TikTok due to the platform's design, which permits them to throw vulnerable comments on blind individuals. These messages also draw on ageist sexism when discussing women’s age and also introduce disability sexualization, portraying womanhood and blindness as things to use. 

While the sexual harassment was serious, there were more invasive comments targeting participants' gender and disability. T37 highlights another aspect of the harassment and entitlement script in which men frame the women’s blindness as “lowered standards” and make themselves seen as desirable:

\begin{quote}
    \textit{Here's a fun one about me being a woman. That is just so awful. I'm constantly getting comments like, "Oh, you're blind. You're like the only woman that I stand a chance with."...dudes constantly commenting, the first part is just hitting on me, saying that I'm attractive. And then they continue, what they're trying to say is, "Oh, you can't see. So that must mean even if I'm ugly, I still have a shot." But really, what they're saying is, "Oh, you're blind. So you must have no standards on physical appearance."} [T37]
\end{quote}

Misogynist entitlement that women should accept any attempt, combined with ableist remarks that blind people have no right to access the appearance, together create coercive demands and pressure in dating. TikTok also denotes this narrative that comments with a blending mixture of sexism and ableism when posted online become more visible and louder, which puts pressure on women about their dating preferences. This stereotype combines and intersects sexism and ableism, transforming access needs to put pressure on women sexually. This pattern was observed and noticed throughout the TikTok platform, where women were contacted every day by strangers. Since messaging and DMs are easy to spread on TikTok, the algorithm-driven visibility multiplies and scales these interactions, intensifying the participants' vulnerability than in offline settings. 







\subsection{Encountering Platform-induced Intersectional Marginalization}

Beyond interactions between audiences and creators, TikTok's platform design and algorithmic management can affect the way blind women and LGBTQ+ creators present themselves. Our analysis focuses on two main areas: (a) platform's design (profile/pronouns and creator tools) and (b) algorithmic management (keyword/visual moderation and reporting). Creators have turned towards workarounds, self-silencing, and narrowing their own expression because these systems have continuously flattened or misread a creator's intersectional identity. Many of our participants described these experiences more as an everyday occurrence rather than an exception. We will begin by presenting how creator tools and profile fields have failed to support blind gender/sexual expression, and then how moderation systems have suppressed specific words and visuals that are used by creators who identify as blind and women/LGBTQ+. The first analysis covers design-level constraints in makeup/effects tools and pronoun/identity fields within TikTok.

\subsubsection{Experiencing Discriminative Platform Design}

Almost all of the participants (N=36) have encountered design gaps like unlabeled effects and binary pronoun menus, which make it hard for blind women and nonbinary/LGBTQ+ creators to express gendered aesthetics and identity. In addition, creators can't safely experiment with looks because TikTok's VoiceOver and effects menus lack description labels. Profile fields on the platform misgender nonbinary identities by excluding pronouns like they/them. For blind women and nonbinary/LGBTQ+ creators, makeup and pronouns are not just cosmetic details; they are how a creator expresses their gender and sexual identity. 

T14, a 52-year-old woman who enjoyed watching female BlindTokers doing makeup, shared her cosmetic-related experiences on TikTok, and how inaccessible editing tools constrain gendered aesthetic play.

\begin{quote}\textit{For blind people, (TikTok) it's just a matter of certain areas not reading and not labeling...If you want to do (video) effects, it may have just like different "makeups" and "smiles." If it gives little detail in your VoiceOver that describes, this will paint your lips red, so that person will know what kind of effect it really is. (It) would help us maybe do more things with it (editing).... if you look at a lot of the blind people, most of them are basic. [T14]} \end{quote}

This is an example of an accessibility metadata gap where T14 mentions how the names of effects like "'makeups' and 'smiles'" aren't read by the VoiceOver feature. Without descriptive labels (e.g., 'paints your lips red'), blind creators can't preview the effects or know how the effects will change their video, thus discouraging the use of complex edits altogether. For participants, makeup can be understood as a gendered practice, so inaccessible tools limit their gender expressions. Thus, creator tools restrict how intersectional identity is experimented with because the platform's UI relies on visual affordances over textual descriptions.

Design constraints are also in identity fields, which can limit gender options and push creators towards fragile workarounds for disclosure. T18 shows what happens when the platform only offers a binary menu for gender.

\begin{quote}
    \textit{On TikTok, there wasn't like a they/them thing, so I just go along with the she/her thing. I do hint on TikTok that I am part of the LGBTQ+ community — I have it on my bio, and I have a rainbow flag in my name. But my siblings noticed that, and they have been threatening me, like, "You have to take it off. If you don't do this, we won't help you with that." And I don’t take it off. My oldest sister was like, "You cannot be blind and LGBTQ at the same time." I'm like, wait, what? How is that? So can I choose? Can I not be blind and be part of the LGBTQ community, or vice versa?} [T18]
\end{quote}

TikTok's interface lacks a they/them pronoun option, which forces creators to use she/her or he/him pronouns that misgender them by design. T18 puts a rainbow emoji and bio text, visual hacks as a workaround to signal queerness. This practice also puts queer visibility on full display, and because blindness is also in play, the combination can make it make signals harder for creators to perceive and easier for others to police. Because of this increased visibility, it can also lead to family surveillance and coercion, as shown in T18's family members' comments. This experience restaged their encounters with control and denial online. Emoji and name hacks are also access-fragile because screen readers may render them inconsistently, so the fix can introduce new barriers. This hack may also result in misclassification and risky disclosure, neither of which respects their intersectional identity. TikTok's platform design makes intersectional identity hard to enter and unevenly legible.


\subsubsection{Facing Algorithmic Management}

Some participants (N=15) also described how TikTok's algorithmic management system determines which intersectional expressions, keyword filters, visual heuristics, and user reporting circulate. We highlight: (a) keyword false positives on identity terms, and (b) visual/modesty policing amplified by reporting. T15 describes how language centered around disability and queerness gets flagged.

\begin{quote}
    \textit{I will usually avoid certain words in my videos. And for example, like the word blind sometimes gets flagged as like bullying or harassment. Calling someone blind, I guess? For example, the word queer on TikTok is something that often violates community guidelines, because a lot of people use the word as an insult and in a derogatory way. But it's also a reclaimed word that a lot of people in the queer community use as a label for themselves.} [T15]
\end{quote}

This is an example of context-blind keyword filtering where terms like "blind" and "queer" are treated as harassment language regardless of speaker or intent. By treating reclaimed/self-descriptive terms in the same way as slurs, moderation down-ranks or removes the very language that creators need. Because both disability and queer labels are risky, blind nonbinary/LGBTQ+ creators face a double bind: it's hard to name or use either, let alone both. As a result, creators adopt code words/euphemisms or avoid these terms entirely, which are forms of self-censorship. This could result in reduced visibility and constrained self-expression. In other words, keyword sensitivity without context suppresses intersectional self-reference. Another mechanism, which involves gendered dressing norms, targets visuals and invites mass reporting. For instance, T10 shared:

\begin{quote}
    \textit{If I posted a video right now, where it looked like I was not dressed well or something, somebody to report it, now we get flagged, even though you can't see anything. It's very odd, and I don't know how they decide the algorithm.} [T10]
\end{quote}

Here, visual/nudity heuristics and user reporting flag perceived immodesty "even though you can't see anything." This happens because automated cues and low report thresholds treat ambiguous dress as a violation and produce false positives. For blind women in particular, this practice is a result of a combination of sexist modesty norms and ableist scrutiny of how a blind woman should look, which makes them easier targets for flags. These risks feed back into a creator's act of self-discipline, but flags can persist despite their compliance. The removals and down-ranking limit a creator's reach and discourage them from expressing themselves in the future, otherwise known as chill future expression. Keyword and visual moderation make intersectional identity hard to say and hard to show.



\subsection{Dealing with Intersectional Marginalization on TikTok}

\subsubsection{Internalizing by Self-discipline}

A significant number of participants (N=14) reported changing their behavior, clothing, voices, and content as strategies to avoid risks on TikTok. This act is called \textit{internalizing by self-discipline}: forcing creators to set their own rules and constraints to avoid harassment or shadowbanning. We will discuss the two main moves that were recorded: (a) changing their clothing to appease the "sighted gaze," and (b) changing their tone and persona to avoid harassment. We will begin by discussing how creators change their wardrobe and manage their clothing to manage the perceptions of the sighted.

Many women and queer creators reported actively altering their clothing to ensure coverage on camera. The participants felt the need to make these changes to avoid lewd speculation, especially since, because of their blindness, they're unable to see how they appear to their audiences on the screen. T10 reports following a wardrobe rule to make sure that her body is covered on camera.

\begin{quote}
    \textit{I try to make sure I'm wearing something on videos that has straps or sleeves. So you can see it... they definitely got me to modify my behavior a little bit.} [T10]
\end{quote}

This is an example of anticipatory self-surveillance where T10 makes a pre-emptive effort to satisfy a sighted 'modesty test' to avoid any sexualized readings and comment disputes such as choosing visible straps or sleeves, so you can see it.' This practice comes from the combination of sexism's modesty policing and the ableist suspicion that blindness affects self-presentation. This narrows gender and queer expression to make sure she isn't misread or hyper-sexualized. Because TikTok is a visual platform with public feedback, this "gaze" is continuous, and the algorithm can continue to bring in new viewers and strangers. As a result, wardrobe constraints become an act of internalized self-discipline that brings offline modesty norms into their digital space, and self-expression is through gender/queer styling choices. 

Some creators also self-discipline by adopting a 'presenter' stance, by changing their voice and persona. This approach involves a calibrated tone that has scripted politeness and matter-of-fact corrections to deflect provocation without escalating it. T39 describes how this tone and routine are needed to validate their capabilities.

\begin{quote}
    \textit{I've always struggled really putting up the real personal side of myself across to everyone. And I think I act more as a presenter than just myself on TikTok, I think that avoids getting negative comments... I have a few comments: "How can you even read the comments if you can't see?" I normally just comment back a very matter-of-fact, just saying I can zoom in. I never attack anyone. I'm very kind and formative and just try to hold my tongue.} [T39]
\end{quote}

T39 notes how she acts more as a 'presenter' to avoid getting negative comments. T39 uses impression management: scripted politeness and matter-of-fact replies, like commenting back, "I can zoom in," to diffuse provocation while educating skeptics. This practice comes from the combination of gendered expectations of niceness and queer respectability pressures, and intersects with ableist competency trials with comments like "how can you even read?" that require creators to continuously demonstrate capability. Because TikTok comments are public and amplifiable, it normalizes these repeated "tests" and reaches wider audiences, further forcing creators to remain calm. The result is that the presenter persona protects the creator's ability to reach more viewers, but adds to emotional labor and sacrifices authenticity. This is a form of internalized self-discipline that mirrors offline tone policing.

Creators internalize the platform's gaze by disciplining their bodies and their personas. These practices come from the combination of gender/sexuality norms and ableist scrutiny, which creates unique constraints on how blind LGBTQ+ creators can appear and behave. Offline tone and modesty policy are re-encountered and intensified on online platforms. This strategy allows creators to broaden their reach, but they have to trade creative energy for self-monitoring and emotional labor.


\subsubsection{Overcoming with Community Support}

Most participants (N=30) highlighted how mutual support and relying on each other mitigate harm and strengthen the collective ties. We frame overcoming through community support, where shared advocacy, exchange of practical strategies, and mutual care convert creators from isolated individuals into collective actors. Especially for blind mothers who identify as LGBTQ+, these ties push back against ableist and sexist doubts while upholding queer presence. We highlight three forms of support: (a) solidarity across movements, (b) everyday modeling and practical knowledge, and (c) group formation beyond TikTok. Participants described community ties as crucial and vital to sustain engagement on TikTok. T03 showed their involvement by posting in explicit support to multiple overlapping causes.  

  \begin{quote}
    \textit{Like fighting against racism, or the rights for LGBTQ+, and also fighting for disability rights. Last year, I posted about fighting against Asian hate.} [T03]
\end{quote}

The quote by T03 represents cross-movement solidarity, whereby promoting everyone’s struggles to the audience increases support, and it transforms individual posting into a collective form of activism. Solidarity by linking race, sexuality/gender, and disability affirms that these are all overlapping identities rather than being distinct. TikTok’s design features, including an algorithmic feed and hashtag syste,m enable coalition-related posts to surface, which spread “stop Asian hate” related posts and amplify campaigns related to LGBTQ+ rights and disability equity. The impact is a sense of collective effectiveness where participants feel supported and have the power to raise their voices to express their views, which makes them feel empowered and also encourages engagement. It is a community through which resilience is tied as a collective strength rather than individual effort. By connecting anti-Asian hate with LGBTQ+ rights and disability advocacy, T03 channels solidarity and strengthens the marginalized voices across. 

Some participants highlighted how simple strategies and role models rendered the stigmatized roles, like blind parenting by mothers, something achievable and more imaginable. T21 explained how they learnt useful and practical solid strategies from blind mothers on TikTok.

 \begin{quote}
    \textit{...But there are other (blind) moms; they've all just been able to adapt to being parents...some of the tips that I've come across that work are like having your kids wear neon colors or bright colors, like labeling and knowing your way around the kitchen and making your kitchen adaptable to you... And I've found that so comforting that I could one day be a parent. That's what TikTok has helped me with.} [T21]
\end{quote}

This reflects peer learning, social support and shared usefulness through which creators demonstrate everyday adaptations for instance, neon clothing, labeling or even accessible kitchen setups that others can use in their own lives. In this context, gendered pressures of ideal motherhood collide with ableist doubts about the abilities of blind mothers, but practical tips provide evidence that these roles are achievable. The guidance is both useful and provides reassurance, instilling that parenthood is attainable and allows people to believe in their own potential, which undermines stigma. TikTok’s brief visual clips allow strategies and practical tips to watch, replicate and also circulate across the community. Community demonstrations and modeling within the community reframes the possibilities and opportunities by motivating individuals to stay active even when under constant judgment or scrutiny. TikTok serves as a platform to circulate everyday knowledge and ensures the flow of practical tips. T21 exemplifies mutual aid in practice, where practical tips, everyday hacks, and role modeling transform doubt into a sense of capability. 

TikTok often led to making connections and community ties beyond the network, which grew into lasting networks of mutual support. T13 described how a single post on TikTok sparked the creation of a mother's support group or community on the platform. 

 \begin{quote}
    \textit{TikTok has made a huge difference. Because I get to hear other people's stories and know that I'm not the only one in the world with blindness, I pull from TikTok. We actually have a moms group with visual impairments that goes on a trip every year. We found each other because a girl posted a video on TikTok saying she was going to start a support group on WhatsApp for moms with visual impairments or blindness. She said, "If you want to join, just comment below." A lot of moms commented, and that’s how she created the WhatsApp chat.} [T13]
\end{quote}

This process operates through mobilizing networks through which an individual’s video call invitation channels responses into a WhatsApp group. The WhatsApp group represents and discusses the intersection of motherhood and disability: the dual pressure on blind mothers as they navigate their gendered caregiving responsibilities and also face challenges regarding disability. TikTok’s visibility makes it easier to find one another, but its design does not allow for group management; however, WhatsApp offers sustained engagement and dialogue. This outcome has resulted in sustained mutual support, through daily conversations and yearly meet-ups that turns isolation into shared community care. The WhatsApp group, which is beyond the mainstream venues, reflects how offline exclusion necessitates the needs for initiation of counterpublics on TikTok. Community building turns temporary visibility into lasting stable structures of mutual support and development. T13 illustrates how TikTok’s discovery features lead to engagement in off-platform groups, which gives rise to a network of care. 

T13’s narratives reflect both empowerment through community and lingering offline judgments: “It’s helped me especially in the BlindTok community... I get to hear other people’s stories and just knowing I’m not the only one.” Her reflections on connecting with other visually impaired moms—“we actually have a moms group with visual impairment that go on a trip every year”—shows how offline stigma was transformed into collective care. Yet, that very group emerged because mainstream spaces doubted or ignored their existence.

Through solidarity posts, everyday modeling and community building creators transform TikTok exposure into shared communal buffering. These forms of support intersect at the crossroads of gender/sexuality norms and ableist doubts validating legitimacy, spreading strategies and reinforcing continuity of voice. Nonetheless, these buffers work alongside the coexisting risks like reporting, misgendering by design, which lead into the following topic of discussion on limits and trade-offs.

\section{Discussion}

\rev{In the findings section, we introduced participants’ intersectional experiences as blind women and LGBTQ+ TikTokers. These experiences include re-encountering offline intersectional marginalization, facing platform-induced marginalization, and navigating intersectional marginalization on TikTok. Together, they span most aspects of participants’ everyday engagement with the platform. To capture the essence of these encounters, we examined their experiences through an infrastructural lens and proposed design implications for future platform development. It is important to note that participants primarily referred to smartphone screen readers when discussing TikTok’s accessibility. Therefore, our design recommendations focus on application features most relevant to screen-reader users. Because blind users rely on diverse assistive technologies—and not all use screen readers—we acknowledge the limitations of this scope.}

\subsection{An Infrastructure Look at Intersectional Experiences on Platforms}

\rev{HCI researchers have been paying constant attention to experiences of marginalized communities, including not only the marginalization experiences caused by social interactions \cite{kojah_silencing_2025} but also how such marginalization are extended to algorithmic platform contexts \cite{kojah_silencing_2025,namvarpour_ai-induced_2025}. Recently, there is a growing line of research focuses on intersectional experiences, such as experiences of trans people of color \cite{modi_finding_2025} and disable women sex workers \cite{rodolitz_as_2025}.} 

\rev{With the current study of blind and women/LGBTQ+ TkTokers, we join this line of work and discuss our findings through the lens of infrastructure studies. We unpack what intersectional experience is, why such experiences form, and how they can be harmful. As introduced in the related work section, infrastructure refers to large-scale sociotechnical systems, where categories, standards, and maintenance in everyday practices are shaped by installed bases. Infrastructures can be highly political. The classification, through which the installed bases are created, could also misrecognize people and practices, creating imbrication (intertwined systems influencing each other) \cite{singh_margins_2017} and residuals (excluded cases infrastructure cannot accommodate) \cite{feinberg_story_2014,ahmed_residual_2015}. By revisiting the notion of infrastructure from an intersectionality perspective, we echo the recent all for more inclusive infrastructure \cite{kirks-cler_towards_2025,lyu_systematic_2025}}

First, we argue that BlindTokers’ intersectional experiences are not isolated incidents; rather, they are infrastructural, woven through multiple, interdependent layers of social life. Infrastructure scholars foreground interconnected components and relational properties that come together for particular purposes \cite{star_steps_1996,lyu_cultural_2022,lyu_reconnecting_2023}. Following this perspective, we frame these experiences as infrastructural. In our study, BlindTokers’ intersectional experiences span online/offline, social/physical, and stranger/intimate contexts. Participants reported identity denial (claims that one cannot be both LGBTQ+ and blind), competence questioning (e.g., doubts about blind women’s capacity for parenthood), and sexual harassment (sexualized or suggestive comments). These dynamics restage offline harms experienced in families, workplaces, and public spaces. Moreover, TikTok’s scale and reach can amplify these dynamics: rather than serving as a reliably safe space, online practices and interactions are refracted back into intimate offline settings and can cause further harm (e.g., identity disclosure on TikTok being discovered by an unsupportive family). These experiences extend beyond identity presentation to directly affect participants’ well-being, mental health, safety, and values. These dynamics are entangled, reflecting systemic discrimination across gender, sexuality, and disability. For instance, identity denial, competence questioning, and harassment often co-occur and cascade across settings, reinforcing one another and intensifying harm.

Second, we analyze how these experiences stem from infrastructural processes. Prior work on social media and justice \cite{moitra_parsing_2021} describes a procedural sequence: classification, imbrication and residuality, and finally torque. We adopt this lens to show how infrastructural processes produce intersectional harms. On TikTok, users and content are governed by algorithms that rely on classification to compress multiplicity. These algorithms constitute the installed base of the infrastructure \cite{star_steps_1996}. When classification schemes fail to fit the world, imbrication emerges—overlapping or layered categories force people’s complex identities into partially compatible slots. Confronted with such imbrication, participants find their multi-faceted identities reduced or substituted; they must, in effect, select a single, more legible category to “fit” the installed base. Otherwise, they are residualized \cite{bowker_sorting_1999}, pushed to the margins as “leftover” cases that the system cannot readily accommodate. Finally, struggling with residuality produces torque: routines are disrupted, identities are misrecognized, and participants bear ongoing social and administrative burdens as they navigate misfit classifications. In our case, socio-technical platforms operate as infrastructures that normalize classification through prevailing social norms and algorithmic rules, channeling people (1) from multiple marginalizations to a single, “acceptable” marginalization (“you can’t be both”), and (2) from one marginalization to another (“choose the bigger” or more socially legible one).

Third, we argue that infrastructural bias can lead to infrastructural violence. Infrastructural violence \cite{rodgers_infrastructural_2012} refers to the ways infrastructures produce, sustain, or intensify harm and suffering in everyday life. Rodgers and O’Neil contend that such violence can result from infrastructural design, function, and breakdown, perpetuating inequality, exclusion, and harm in modern societies. In technology-mediated contexts, Li and Nardi \cite{li_there_2021} describe epistemic, emotional, and interpersonal forms of violence. In our study, participants first confront epistemic harms (denial or erasure of their existence), then emotional harms (exposure to harsh or demeaning commentary), and finally interpersonal harms (strain and rupture in intimate relationships, such as family).

Taken together, the infrastructural perspective (1) highlights the interconnectedness of intersectional experiences and how they are generated and regenerated; (2) extends beyond identity multiplicity to analyze how platform infrastructures cause and pattern these experiences; and (3) reveals infrastructural violence as a downstream consequence, specifying the harms participants face. This perspective offers a rich analytic approach, units of analysis, and implications for future theoretical exploration of identity on algorithmic platforms.

\subsection{Facilitating Blind Women/LGBTQ+ TikTokers' Infrastructuring Work with Platform Design}

In the face of the aforementioned infrastructuralized intersectional experiences, users work on maintenance and repair to keep delicate infrastructure functioning. This includes daily fixes, the creation of ad-hoc standards, and efforts toward community maintenance, all of which sustain vulnerable technologies. In our study, we portray the coping strategies of participants as "infrastructuring work \cite{pipek_infrastructuring_2009}", the continuous, situated maintenance required to keep creative practice possible for blind women/LGBTQ+ creators. This maintenance involves reclassifying, remediating, and rerouting these systems solely to enable their identity to be entered, perceived, and circulated on TikTok. \rev{Prior work has shown how other TikTokers with disabilities, such as people with attention-deficit/hyperactivity disorder (ADHD) \cite{simpson_hey_2023}, conduct infrastructuring work with creative labor to fix accessibility issues.} From our data, we see this infrastructure appearing as schema signaling, context provisioning, creative accessibility hacks, mutual-aid circulation, and off-platform anchoring. Reframing these “coping” strategies as infrastructuring shifts the responsibility from individual resilience to system repair and highlights where design is responsible for a burden. Next, we propose design solutions that reduce the need for self-discipline while interrupting the pipeline from classification to torque.

\begin{enumerate}
\item Identity fields and markers should be treated as first-class schema, and not as user hacks \cite{keyes_misgendering_2018,scheuerman_how_2019}. This would mean adding custom and multi-pronouns with text-to-speech labels and audience-scoped visibility, and providing machine-readable equivalents to emoji markers. This interrupts the classification collapse that suggests, “you can’t be both,” and reduces the epistemic and interpersonal harms of misgendering and family coercion. The success of this can be tracked by reductions in misgendered profiles and the forced use of workarounds by blind women/LGBTQ+ creators.

\item Moderation systems should distinguish differences between self-description and attack and actively detect brigading \cite{haimson_disproportionate_2021, starbird_disinformation_2019}. This would require using identity whitelists and signals for speakers and targets, clustering reports, and raising thresholds in likely brigades. It requires human review for cases at the intersection of disability and gender, and a restoration of content reach upon reversal. This approach breaks the imbrication between mob reports and visual or keyword-based false positives, and lowers emotional and interpersonal harms for blind women/LGBTQ+ creators. Success can be measured by publishing false-positive rates and post-appeal visibility restitution statistics by disability, gender, and sexuality.

\item Platforms should expand their notion of accessibility to cover both content creation and content watching \cite{huh_avscript_2023, zhang_screen_2021}. This requires implementing text-to-speech–readable effect descriptions, audio previews, and full VoiceOver navigation within the editors. This reduces residuality by eliminating the need to be “basic by necessity” and reduces epistemic harm by enabling blind women/LGBTQ+ creators to show gendered aesthetics. Its success can be measured by tracking the edit complexity achieved with screen readers and surveying creators on their perceived creative range.

\item Platforms should also shift routine safety work from people to presets \cite{wei_theres_2023}. This can be achieved by providing creators with more controls, such as follow-only comments on identity posts, first-time commenter approvals, and DM throttles. These should also include one-tap auto-replies, like “I use magnification/zoom.” This reduces the need for self-discipline and lowers emotional labor for blind women/LGBTQ+ creators because it disrupts the imbrication of sexist and ableist scripts with open contact. Success can be tracked by measuring reductions in time spent moderating and the number of repeat harassment chains.

\item TikTok must legitimize and protect counter-infrastructures built by blind women/LGBTQ+ creators \cite{nguyen_it_2025,melder_blocklist_2025}. This means providing primitives that enable a seamless flow from post to group, including a consent gate, moderator roles, and safety defaults. It also means building searchable, text-to-speech–structured tip repositories with provenance. This intervention stabilizes mutual-aid circulation and lowers interpersonal harm by providing access to moderated, safe spaces. Success can be measured by tracking creator retention within these creator groups and the incident rate per member.

\item Platforms should make risks and decisions visible to creators through accountability dashboards \cite{liang_is_2025, wu_negotiating_2024}. These dashboards could include per-post moderation traces, “likely-to-flag” alerts, and weekly chilling-effect metrics by disability, gender, and sexuality. This intervention targets the epistemic harm of opaque decisions and enables an audit of disparate chilling effects on blind women/LGBTQ+ creators. Its impact can be measured by a difference-in-differences analysis of content removals and down-ranking after the dashboard’s launch.

\end{enumerate}

To conclude, seeing participants’ strategies as a form of infrastructuring explains how their expression survives despite biased classification and brittle tools. Our design implications institutionalize this labor and shift the burden from privatized maintenance to system-level support that interrupts the classification-to-torque pipeline. This approach reduces epistemic, emotional, and interpersonal harms while expanding creative and communal horizons for blind women/LGBTQ+ creators.

\section{Limitations}

\rev{This study investigates blind women and LGBTQ+ TikTokers’ intersectional experiences. We examined their encounters across both online and offline social contexts and proposed design implications to inform future technologies. That said, we acknowledge several limitations of this work. First, we acknowledge limits related to the research team’s background. Our team consists of sighted, non-disabled researchers; some members are cisgender straight women, and none of us identify as part of LGBTQ+ communities. This absence of blind women and LGBTQ+ researchers may limit our ability to fully capture the nuances of participants’ intersectional experiences. Second, we acknowledge limitations in our sampling method. We recruited participants from a list of TikTokers who used keywords such as “blind” or “visual impairment” in their usernames, bios, or content. However, this approach may have excluded blind TikTokers who do not disclose such information on their accounts. Third, we acknowledge the limited scope of our design implications. Our recommendations focus primarily on the TikTok smartphone application and screen-reader use. As a result, they may not fully address the needs of blind users who rely on other accessibility tools, such as magnifiers. We call attention to these limitations and encourage future researchers to join us in addressing them through more inclusive, expansive, and collaborative work.}
\section{Conclusion}
The HCI community has been paying increasing attention to social media platforms as they are important venues for identity expression. In this paper, we studied a population, blind people who are also women and/or LGBTQ+, on TikTok. We interviewed 42 BlindTokers and identified their intersectional experiences mediated by TikTok. We found that BlindTokers' intersectional marginalization is infrastructural and use this infrastructure perspective to recognize participants' infrastructuring work to address the problems. We also provide implications for future social media design and policy-making.

\begin{acks}
Thanks for the reviewers' comments. 
\end{acks}

\bibliographystyle{ACM-Reference-Format}
\bibliography{Reference/Yao}

\end{document}